# SENTINEL: Securing Indoor Localization against Adversarial Attacks with Capsule Neural Networks

Danish Gufran, Pooja Anandathirtha and Sudeep Pasricha
Department of Electrical and Computer Engineering,
Colorado State University, Fort Collins, CO, USA
{Danish.Gufran, Pooja.Anandathirtha , Sudeep}@colostate.edu

*Abstract*— **With the increasing demand for edge device powered location-based services in indoor environments, Wi-Fi received signal strength (RSS) fingerprinting has become popular, given the unavailability of GPS indoors. However, achieving robust and efficient indoor localization faces several challenges, due to RSS fluctuations from dynamic changes in indoor environments and heterogeneity of edge devices, leading to diminished localization accuracy. While advances in machine learning (ML) have shown promise in mitigating these phenomena, it remains an open problem. Additionally, emerging threats from adversarial attacks on ML-enhanced indoor localization systems, especially those introduced by malicious or rogue access points (APs), can deceive ML models to further increase localization errors. To address these challenges, we present SENTINEL, a novel embedded ML framework utilizing modified capsule neural networks to bolster the resilience of indoor localization solutions against adversarial attacks, device heterogeneity, and dynamic RSS fluctuations. We also introduce *RSSRogueLoc*, a novel dataset capturing the effects of rogue APs from several real-world indoor environments. Experimental evaluations demonstrate that SENTINEL achieves significant improvements, with up to 3.5× reduction in mean error and 3.4× reduction in worst-case error compared to state-of-the-art frameworks using simulated adversarial attacks. SENTINEL also achieves improvements of up to 2.8× in mean error and 2.7× in worst-case error compared to state-of-the-art frameworks when evaluated with the real-world *RSSRogueLoc* dataset.**

*Index terms*— **Adversarial attacks, rogue access points, evil twin attacks, man-in-the-middle attacks, adversarial training, device heterogeneity, wi-fi RSS fingerprinting, capsule neural networks.**

## I. INTRODUCTION

IN recent years, indoor localization has gained attention for its versatile applications across several industries, such as healthcare, asset tracking, smart homes, location-based advertising, and much more [1]. The ability to pinpoint the exact location of edge devices within indoor settings has the potential to revolutionize these industries and elevate user experiences significantly. Hence, technology giants such as Apple, Google, Meta, and Microsoft are making substantial investments in indoor localization research to improve the accuracy and reliability of indoor location-based services [2]. However, achieving high-precision indoor localization remains a formidable challenge due to the inherent complexities and dynamic nature of indoor environments.

Traditional navigation systems, such as the global positioning system (GPS), have found widespread adoption in popular tools such as Google Maps, Apple Maps, and Waze, mainly owing to their commendable localization accuracies in outdoor settings. However, the dependence of GPS on satellite signals and clear sky visibility poses a significant limitation, rendering this approach ineffective for indoor use [3]. In response to this challenge, researchers have shifted their attention to alternate wireless infrastructures that could be a better fit for localization across indoor spaces, such as Wi-Fi, Bluetooth, and Zigbee. Among these alternatives, Wi-Fi-based localization systems utilizing received signal strength (RSS) have gained significant traction [1]-[4]. This surge in popularity for this solution is attributed to the ubiquitous availability of Wi-Fi in indoor spaces and the capability of modern edge devices to capture Wi-Fi RSS, making it a viable option for indoor localization [4].

Wi-Fi RSS is obtained by measuring the signal strength of nearby Wi-Fi routers or access points (APs) via edge devices. This captured RSS data can be used to estimate the current indoor location of an edge device. As the edge device moves, it periodically captures new RSS measurements, reflecting the edge device's mobility. Leveraging this changing RSS data, many techniques have been proposed for accurate indoor localization, with geometric model-based [5] and fingerprinting model-based [4], [6] approaches emerging prominently. Geometric models utilize propagation methods such as trilateration [7] and triangulation [8] to pinpoint an edge device's location. However, these solutions are prone to inaccuracies as they are particularly sensitive to RSS fluctuations caused by dynamic changes and complexities within indoor environments. On the other hand, fingerprinting model-based systems eschew propagation methods by creating a database of Wi-Fi RSS patterns ("fingerprints") of visible Wi-Fi APs collected throughout the indoor space to estimate location. Fingerprinting models have been shown to exhibit greater resilience to RSS fluctuations, demonstrating higher accuracies than geometric methods [4], [9].

Fingerprinting-based localization solutions comprise of two distinct phases: an offline phase and an online phase. During the offline phase, Wi-Fi RSS fingerprints are systematically captured across multiple reference points (RPs) within a building floorplan. Typically, multiple fingerprints are recorded per RP to accommodate data variability that can arise due to RSS fluctuations in the online phase. These fingerprints are then often utilized to train a machine learning (ML) model, enabling it to capture underlying patterns and features within the collected RSS fingerprints [10]. Once trained, this ML model is deployed on the edge device, making it available in the online phase for real-time indoor location predictions.

In the online phase, the RSS fingerprints may exhibit fluctuations due to diverse factors in the indoor environments. These factors include signal attenuation, reflections from objects, human interference, and multipath fading, which can introduce fluctuations in the collected RSS fingerprints [11]. Furthermore, edge device heterogeneity exacerbates this issue. Even among edge devices utilizing the same Wi-Fi chipset (from the same manufacturer), differences in hardware, software, antenna configurations, and firmware settings can introduce fluctuations in RSS fingerprints [11]. As a result, training an ML model can be challenging as heterogeneous and noisy RSS can result in poor generalization and result in inaccurate location predictions. Priors works have shown up to a 41% reduction in location accuracy due to these factors [12]. Additionally, the often-overlooked factor of adversarial attacks can not only perturb the RSS fingerprints (thereby introducing stronger fluctuations) but also compromise the accuracy and effectiveness of localization with the edge device, emphasizing the need for more robust and secure localization systems.

Adversarial attacks can mislead popular ML models, including state-of-the-art deep learning (DL) algorithms that have been shown to be vulnerable to adversarial examples. The authors of [13] verified the discovery by misleading the popular GoogLeNet [14] model with adversarial examples. Similarly, ML based indoor localization systems also face the threat of adversarial attacks. The presence of malicious (or rogue) APs in the building floorplan can be used to create adversarial attacks by mimicking a legitimate AP and broadcasting erroneous RSS values. In Fig. 1, we illustrate the detrimental impact of the presence of rogue APs on three popular ML-based indoor localization solutions based on K-Nearest Neighbors (KNN) [15], Gaussian Process Classifier (GPC) [16], and Deep Neural Networks (DNN) [17]. This experiment was conducted on an indoor path in a building measuring 55 meters in length containing 55 RPs (1 RP per meter), with up to 203 visible APs (per RP). The experiment incorporated the popular fast gradient sign method (FGSM) [30] technique to simulate the presence of rogue APs, resulting in significantly increased indoor localization errors, with average error increases of 3.33× for KNN, 3.0× for GPC, and 5.71× for DNN, highlighting the negative impact of the rogue APs on localization accuracy.

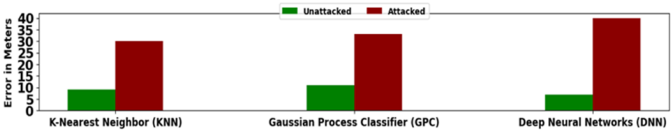

**Fig. 1.** Impact of rogue APs on three popular ML-based indoor localization solutions [15]-[17] from prior work.

To tackle the challenges posed by RSS fluctuations in dynamic indoor environments, edge device heterogeneity, and rogue AP attacks, in this work we introduce SENTINEL, a novel embedded ML framework that employs modified capsule neural networks tailored specifically for indoor localization and rogue AP resilience, offering a more practical, secure, and real-time solution for indoor localization. The major contributions of our SENTINEL framework are:

- We design a novel modified capsule neural network specifically for the RSS fluctuation challenges in indoor localization, tailored to 1) overcome the spatial invariance problem in prior DL-based indoor localization efforts and 2) enable lightweight deployment on edge devices.
- We study the effects of rogue AP attacks and propose an adversarial training setup together with the modified capsule neural network for resilience against adversarial (rogue) AP attacks for the first time in indoor localization.
- We introduce a new Wi-Fi RSS fingerprint dataset called *RSSRogueLoc* [35] that captures AP attacks from rogue APs in real-world indoor environments for the first time.
- We conduct a performance comparison with SENTINEL against state-of-the-art indoor localization solutions, to highlight its effectiveness in the presence of diverse adversarial attacks, edge device heterogeneity, and RSS fluctuations across diverse indoor building paths.

## II. RELATED WORK

Wi-Fi fingerprinting-based indoor localization has gained significant recognition, evident in competitions hosted by industry giants like Microsoft and NIST [2]. Several classical ML-based solutions, such as ones based on the KNN [15] and GPC [16] algorithms have showcased their potential in addressing RSS fluctuations arising from dynamic effects in indoor environments. These fluctuations encompass various factors, including human interference, obstacles, movement of furniture or equipment, variable population density, signal interference, reflections by objects, and shadowing effects [19], [40], [41].

Despite the demonstrated promise of these ML solutions, they often face challenges in maintaining robustness against fluctuations introduced by edge device heterogeneity. The heterogeneity issue arises from differences in Wi-Fi chipsets and noise filtering software employed by different manufacturers of edge devices. As these chipsets and software stacks are crucial for extracting RSS fingerprints [11], [19], the heterogeneity within them introduces additional complexities for traditional ML-based indoor localization systems.

In response to these challenges, researchers have explored the use of more powerful DL algorithms for indoor localization, including DNNLOC [17], MLPLOC [18], LC-DNN [19], CNNLOC [21], SANGRIA [22], ANVIL [23], and TIPS [24]. DNNLOC [17], MLPLOC [18], and LC-DNN [19] employ DNNs along with improved RSS pre-processing methods to enhance feature correlation in the RSS fingerprints. CNNLOC [21] proposes a modified convolutional neural network (CNN), to improve on these efforts by enhancing the model's ability to capture relevant features in the RSS fingerprints. SANGRIA [22] employs DNN based autoencoders while ANVIL [23], [42] utilizes attention neural networks, to improve focus on critical input features. TIPS [24] leverages transformer-based encoding of RSS fingerprints for improved resilience against fluctuations introduced by dynamic indoor environments and device heterogeneity. However, these approaches are still significantly impacted by more complex heterogeneity effects in emerging devices and are also susceptible to adversarial attacks, due to the spatial invariance problem in DL algorithms.

Most DL algorithms, particularly CNNs, suffer from the spatial invariance problem where the DL algorithm has a propensity to focus solely on the presence of features in the data while neglecting the precise relative positions of the features

[25]. Alterations in the position of each feature can lead to mispredictions by the DL model. This limitation is illustrated in Fig. 2, where the VGGFace algorithm [26], a CNN-based model, struggles to differentiate between the two faces. In the figure on the left, a normal human face is depicted, while the figure on the right presents an abnormal face with jumbled feature positions. The model assigns the same output classification probability to both cases. The concern regarding feature positions is particularly relevant in the context of RSS fingerprints for indoor localization, where positions of certain features represent crucial information and can be specific to a particular RP. When an edge device moves to a different RP, the positions of these features may undergo changes based on the characteristics of the new RP location. Thus, it is imperative to account for the dynamic nature of feature positions when designing practical indoor localization solutions.

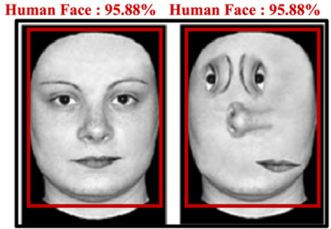

**Fig. 2.** Spatial invariance problem in deep learning algorithms. Both cases are classified as valid human faces by a CNN model.

To address this limitation and enhance feature extraction, researchers have embraced more recent DL algorithms, such as vision transformers (VITAL) [27], [43] and capsule neural networks (EDGELOC) [28] for indoor localization. VITAL [27], uses vision transformers, introduces positional encoding for each feature, aiming to overcome the spatial invariance limitations posed by CNNs. Similarly, EDGELOC [28] uses a simple capsule neural network derived from [38], treating each captured feature as a vector, considering both magnitude and direction of features. These frameworks show the potential to greatly mitigate the effects of dynamic environments and heterogeneity for indoor localization. However, the introduction of adversarial attacks especially arising from rogue APs can not only jumble the feature positions but also introduce new malicious features in the data. Such attacks can easily mislead state-of-the-art localization frameworks and compromise user security.

Adversarial training has emerged as a potential solution to address the challenges from adversarial attacks in ML [29]. Popular solutions typically incorporate a subset of adversarial samples along with the training data to allow robustness in the presence of adversarial attacks during inference. Adversarial samples are generated using several popular adversarial methods out of which the fast gradient signed method (FGSM) [30] has been widely employed to simulate the effects of adversarial attacks, owing to its simplicity. ADVLOC [31] and CALLOC [32] are two recent solutions that incorporate adversarial training, aiming to address the effects of adversarial attacks in indoor localization. Both ADVLOC [31] and CALLOC [32] integrate FGSM samples during training for adversarial resilience. CALLOC additionally employs curriculum learning along with attention neural networks to enhance feature extraction between the original and adversarial samples, to improve overall robustness. Nevertheless, both solutions fall short of addressing the multitude of challenges associated with dynamic environments, heterogeneity, and adversarial attacks concurrently. Additionally, these solutions heavily rely on simulated data for measuring the efficacy of the model's performance against adversarial attacks in the online phase. Their performance in real-world adversarial scenarios has not yet been carefully studied.

After carefully studying the simultaneous challenges of dynamic environments, edge device heterogeneity, adversarial attacks, and lack of real-world adversarial attack data to measure the effectiveness of adversarial resilience in indoor localization, in this work we propose SENTINEL, a novel embedded ML framework that goes beyond state-of-the-art DL solutions to better address the spatial invariance problem and improve robustness using an enhanced capsule neural network with techniques that more comprehensively improve resilience to real-world indoor localization challenges. Another important contribution of our work is the design of a newly curated RSS fingerprint dataset called *RSSRogueLoc* [35] that captures the presence of rogue APs within indoor building paths, to analyze the impact of adversarial attacks on indoor localization frameworks in real-world environments, for the first time.

III. ADVERSARIAL ATTACKS IN INDOOR LOCALIZATION

Adversarial attacks involve deliberately perturbating input data to deceive an underlying ML model [30]. This perturbation typically consists of adding noise to individual data values (datapoints) either by introducing new or malicious features (new datapoints) or disrupting the magnitude and positions of features in the input data. These adversarial perturbations exploit limitations in the manner in which features and patterns are learned by the ML model during training, thereby causing mispredictions with the ML model [30].

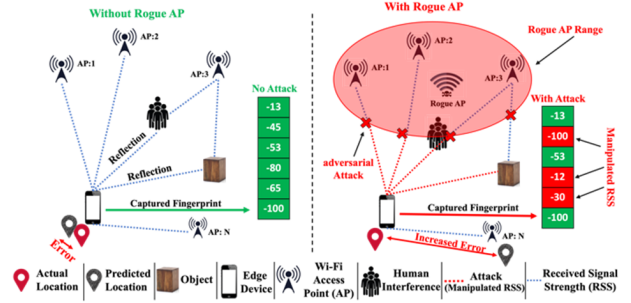

**Fig. 3.** RSS fluctuations in indoor environment depicting real-world scenarios with and without the presence of rogue APs.

In the context of indoor localization, Wi-Fi RSS fingerprints are measured in decibels referenced to one milliwatt (dBm) and typically range from -100 dBm (weak signal) to 0 dBm (strong signal). These fingerprints are very susceptible to fluctuations due to dynamic indoor environments and edge device heterogeneity, and perturbations due to adversarial attacks especially in the presence of rogue APs, as shown in Fig. 3. Rogue APs can perturbate specific or all datapoints within an RSS fingerprint. This perturbed data may exhibit features characteristic of a different RP location, leading to increased prediction errors, as shown in Fig. 3.

Rogue APs pose a threat to indoor localization systems by introducing deliberate perturbations through two distinct pathways: the transmitter side (involving APs) and the channel side (within the space between the AP and edge device).

- **Transmitter side:** This attack is executed from the transmitter side, specifically on the APs deployed in the indoor environment. The attack targets a legitimate AP in the environment, attempting to infect it with malicious data (malware). Once successful, the resulting rogue AP gains complete control over the legitimate AP, compromising the security of any operations performed by the legitimate AP. This poses a significant security risk, as the rogue AP can now manipulate RSS, leading to an increase in localization errors. This attack can compromise the robustness of the indoor localization solution in that environment.

- **Channel side:** This attack is executed from the channel side, specifically within the spatial domain between a legitimate AP and the edge device. The rogue AP monitors communication between the legitimate AP and edge devices and introduces carefully calibrated interference with the signals traveling through this space. Once successful, the rogue AP can manipulate the RSS captured by the edge device, that may mimic the characteristics of a different RP location. This manipulation compromises the robustness of the indoor localization solution, as the altered RSS can lead to increase in localization errors.

*A. Rogue AP Attack Implementation*

Rogue APs possess the capability to execute a variety of attacks. In the context of indoor localization, we delve into the practical implementation of these attacks, focusing on the transmitter side (evil twin attacks) and the channel side (man-in-the-middle attacks) within the indoor localization system. Notably, these attacks can be launched with minimal information about the target system, rendering them as grey box attacks. The nature of grey box attacks makes rogue APs an attractive choice for adversaries, as they do not require comprehensive knowledge of the indoor localization system. Unlike traditional white box attacks requiring complete access to privileged information, such as ML model architecture, localization framework policies, building layouts, and access to edge devices, grey box attacks operate with partial knowledge. This characteristic transforms rogue AP implementation into a more plug-and-play system for executing adversarial attacks. We next describe the two types of rogue AP attacks, illustrating their underlying methods and potential consequences.

- **Evil twin attacks:** This transmitter side rogue AP attack involves the creation of a malicious wireless network that mimics a legitimate one. The rogue AP utilizes malware to infect a legitimate AP, allowing it to gather critical information such as the SSID (service set identifier), MACID (media access control identifier), and other network parameters [36]. By replicating these parameters, the rogue AP tricks edge devices into connecting to it, masquerading as an authentic AP. Fig. 4 demonstrates the implementation of the evil twin attack, which is explored for the first time in the context of indoor localization, as part of this work. The rogue AP initiates the attack by targeting a legitimate Wi-Fi AP, mimicking its network parameters, and simultaneously blocking all communications from the legitimate Wi-Fi AP. Subsequently, the rogue AP broadcasts its own malicious Wi-Fi network (masquerading the authentic AP), that can inject malicious features into the RSS fingerprint collected by the edge device. These malicious features have the potential to falsify the edge device's perceived location, making it appear in a different location. This compromise in location information poses a severe threat to the entire indoor localization system.

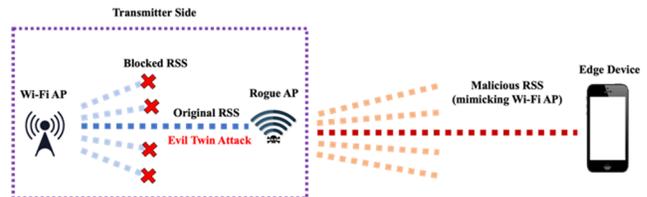
**Fig. 4.** Evil twin attack during indoor localization.

- **Man-in-the-middle attacks:** This channel side rogue AP attack employs ARP (address resolution protocol) spoofing techniques to intercept communication between the legitimate Wi-Fi AP and the edge devices [37]. Operating within the spatial domain between the AP and the edge device, the rogue AP positions itself as an intermediary, intercepting signals transmitted between the legitimate AP and the edge device. Unlike direct communication, the man-in-the-middle attack allows the rogue AP to inspect, modify, or block the signals before relaying them to their intended destination. This interception provides the adversary with the capability to alter RSS values in real-time, introducing discrepancies in the RSS features captured by the edge device. Fig. 5 demonstrates the implementation of the man-in-the-middle attack for indoor localization.

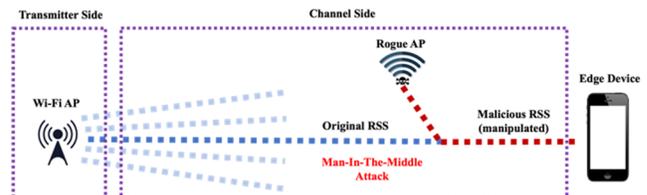
**Fig. 5.** Man-in-the-middle attack during indoor localization.

*B. Adversarial Attack Methods*

Adversarial perturbations, introduced by malicious entities, pose a threat to ML models, particularly in privacy-sensitive domains like indoor localization. The covert nature of these attacks necessitates robust mitigation methods. We identify and focus on three popular adversarial methods in this work: fast gradient sign method (FGSM) [30], projected gradient descent (PGD) [33], and momentum iterative method (MIM) [34]. Given the grey box nature of adversarial attacks (evil twin and man-in-the-middle attacks, discussed above), adversaries exploit minimal information about the localization framework. These methods introduce carefully calibrated perturbations into the RSS fingerprints using the ML model's loss function, making them a practical choice for studying the nuanced effects of adversarial attacks in indoor localization. Each method (FGSM, PGD, MIM) serves a unique purpose in aiding our understanding of adversarial attacks in indoor localization systems. These methods are briefly described below:

- **Fast gradient sign method (FGSM):** FGSM leverages the gradient information of the ML model's loss function with respect to the input data. This method perturbs the original input data by adding a small, controlled perturbations in the direction of the gradient sign. This intentional perturbation systematically alters both the magnitude and positions of features within the input data. In the context of indoor localization, perturbations induce changes in the values within the RSS fingerprint, effectively manipulating the magnitude and positions of features inherent in fingerprint data. Consequently, this perturbation can mislead the ML model by indicating features at a different RP location, thereby increasing errors in location predictions.

$$\eta = \epsilon * sign(\nabla J(\theta, X, Y)) \quad (1)$$

$$X_{Adv} = X + \eta \quad (2)$$

In the equations above, $\eta$ represents the perturbations, $\theta$ represents the parameters of the ML model, and $X$ and $Y$ denote the RSS fingerprint and RP class, respectively. The hyperparameter $\epsilon$ controls the magnitude of the perturbation and $(\nabla J(\theta, X, Y)$ denotes the loss function of the ML model. $X_{Adv}$ is the perturbated RSS data.

- **Projected gradient descent method (PGD):** PGD extends the concepts of FGSM by offering a more sophisticated approach in generating adversarial examples. PGD modifies FGSM by eliminating the sign function in equation (1) and clipping the perturbations between $X$ and $\epsilon$. While FGSM introduces perturbations in a single step, PGD refines the perturbation over multiple iterations $\{X_{Adv\,(0)}, X_{Adv\,(1)},..., X_{Adv\,(N)}, X_{Adv\,(N+1)}\}$.

$$X_{Adv\,(0)} = X \quad (3)$$

$$\eta = Clip_{X,\epsilon}\{\epsilon * \frac{\nabla J(\theta,X,Y)}{L|\nabla J(\theta,X,Y|_2}\} \quad (4)$$

$$X_{Adv\,(N+1)} = X_{Adv\,(N)} + \eta \quad (5)$$

In equation (3), $X$ denotes the original input data and $X_{Adv\,(0)}$ denotes the perturbed adversarial sample at the initial iteration (0). Equation (4) computes perturbations $\eta$ using a clipped function applied to the gradients of the loss function $\nabla J(\theta, X, Y)$ and $L|\nabla J(\theta, X, Y|_2$ represents the squared L2 norm (ridge regularization) of the gradients of the loss function. This normalization step ensures that the perturbation is scaled appropriately, maintaining stability in generating the adversarial sample, while being clipped between $X$ and $\epsilon$ (magnitude of the perturbation). These perturbations are added to $X_{Adv\,(N)}$ iteratively, as shown in equation (5). This iterative refinement process enhances the potency of adversarial samples by introducing a more calibrated manipulation in feature magnitude and positions within the RSS fingerprint data, leading to more potent adversarial samples compared to FGSM.

- **Momentum iterative method (MIM):** MIM further refines the adversarial samples from PGD, by incorporating momentum into the perturbation generation process to enhance the efficiency of the perturbation search.

$$X_{Adv\,(N+1)} = Clip_{X,\epsilon}\{\alpha * X_{Adv\,(N)} + \eta\} \quad (6)$$

The perturbation $\eta$ is calculated using equation (4), similar to the PGD approach. In equation (6), $\alpha$ is applied as momentum to the $X_{Adv\,(N)}$ of the previous iteration, while being clipped between $X$ and $\epsilon$ (magnitude of the perturbation). By incorporating momentum into the perturbation generation process, MIM effectively manipulates RSS features and positions, leading to adversarial samples that induce more significant errors in the localization process, compared to FGSM and PGD. This enhanced perturbation poses substantial challenges to the robustness of indoor localization solutions.

*C. Adversarial Attack Formulation for ML Indoor Localization*

In formulating adversarial attacks for indoor localization systems, we employ the three distinctive methods discussed above: FGSM, PGD, and MIM. Our objective is to generate adversarial data by introducing perturbations that modify the features embedded within an RSS fingerprint. This strategic perturbation of the RSS data is crucial in mirroring potential real-world attack scenarios, where rogue APs manipulate signals, thereby manipulating RSS features to deceive a localization solution. To generate potential real-world adversarial data effectively, we leverage two key parameters:

- **Perturbation strength ($\epsilon$):** This crucial hyperparameter is used in FGSM, PGD, and MIM methods to introduce perturbations to the RSS fingerprints. In generating adversarial samples for indoor localization, we systematically adjust the $\epsilon$ value to encompass various perturbation strengths applicable in real-world scenarios. We vary $\epsilon$ from 0.1 to 0.5 to reflect a practical perturbation scenario tailored for indoor localization [39]. This range is considered acceptable because it strikes a balance between being subtle enough to evade detection and significant enough to effectively test the system's robustness. Smaller values of $\epsilon$ (closer to 0.1) represent minor perturbations that are less likely to be noticed but might not challenge the system's defenses effectively, while larger values (up to 0.5) represent more noticeable perturbations that can more rigorously test the model's resilience. To emulate potential threats, rogue APs strategically deployed in indoor spaces can introduce these perturbations to mislead localization systems.

- **Compromised APs ($\varphi$):** This parameter represents the quantity of legitimate APs that are subject to compromise by the rogue AP within the indoor system. In a typical scenario, rogue APs selectively attack a subset of legitimate APs. We utilize $\varphi$ as a parameter to investigate the impact of the quantity of compromised APs on indoor localization performance. $\varphi$ is set to range from 0 to 100, indicating the percentage of attacked APs, thus covering the spectrum from 0% to 100% of compromised APs. These attacked APs then introduce perturbations defined by the parameter $\epsilon$.

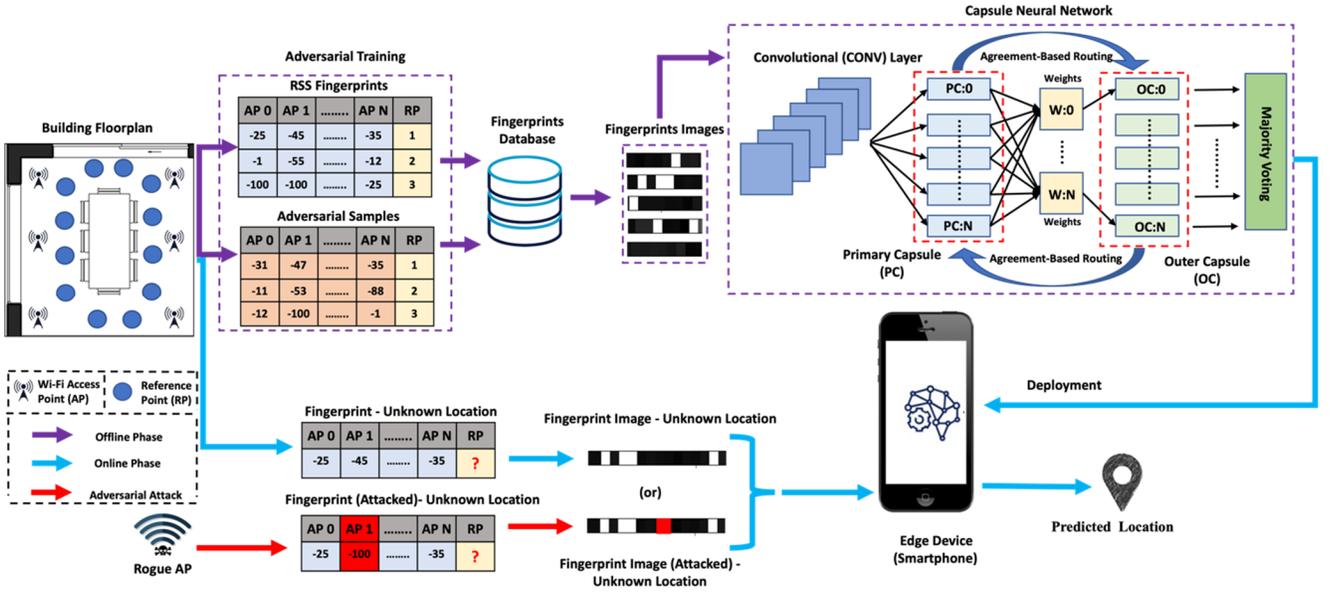

**Fig. 6.** Overview of the SENTINEL framework, including the offline (training) phase and online (inference) phase.

IV. SENTINEL FRAMEWORK: OVERVIEW

The SENTINEL framework consists of three key components: adversarial training, fingerprint image generation, and the capsule neural network, as shown in Fig. 6. The framework initiates in an offline phase, where RSS fingerprints are captured across different RPs within the building floorplan. Multiple fingerprints are collected per RP to effectively capture data variability. These fingerprints are labeled and stored in an RSS fingerprint database, forming the offline training data for the SENTINEL framework. To fortify the framework against adversarial attacks, we employ an adversarial training mechanism (discussed in Section IV.A), which introduces adversarial samples derived from the RSS fingerprint database. This approach not only enhances the framework's resilience towards potential adversarial challenges but also introduces additional variability to the training data, thereby improving the framework's ability to enhance its resilience towards dynamic indoor environments and edge device heterogeneity in the online phase. Post-adversarial training, we transform both original (from RSS fingerprint database) and adversarial fingerprints into fingerprint images using the fingerprint image generation mechanism (discussed in Section IV.B), resulting in greyscale images. These grayscale images encapsulate crucial information about the indoor floorplan. The grayscale images then serve as input to the capsule neural network modified for the task at hand and carefully designed to address the spatial invariance problem in DL. The capsule neural network comprises of five sub-components: convolutional layer (CONV), primary capsule layer (PC), outer capsule layer (OC), an agreement-based routing algorithm, and the majority voting layer (all discussed in Section IV.C).

The domain-specific capsule neural network, once trained, is deployed on edge devices for predictions during the online phase. In the online phase, the edge devices (with the pre-trained ML model), scan for available RSS fingerprints at an unknown RP location. These received fingerprints are inherently susceptible to RSS fluctuations (introduced by dynamic environments and edge device heterogeneity) and potential adversarial attacks (introduced by rogue APs). Nevertheless, the SENTINEL framework is able to achieve resilience against these challenges, as discussed in Section V via sensitivity analysis experiments across devices and building, and comparison with state-of-the-art frameworks.

*A. Adversarial Training Mechanism*

The SENTINEL framework enhances its resilience against adversarial attacks by implementing an adversarial training mechanism. This approach fortifies our capsule neural network by exposing it to a diverse mixture of adversarial and clean RSS examples during the training process. The fundamental concept behind adversarial training is to modify the loss function by incorporating adversarial examples, thereby rendering the capsule neural network resistant to adversarial attacks.

$$\nabla J(\theta, X, Y) = \nabla J(\theta, X, Y) + \nabla J(\theta, X + \eta, Y) \quad (7)$$

In equation (7), $\eta$ represents the perturbation introduced into the input data using different adversarial methods such as FGSM, PGD, and MIM, calculated using the gradients of the loss function (equations (1), (3), and (5)) with respect to the input data. In Section V, we evaluate the performance of various adversarial training methods to assess SENTINEL's efficacy in defending against adversarial attacks in the online phase.

*B. Fingerprint Image Generation*

Post creation of the RSS fingerprint database (with clean + adversarial samples), the fingerprints are transformed into grayscale images to encapsulate crucial information about the indoor floorplan. Initially, the RSS fingerprints are arranged into matrices or tensors, with shape of (H, W), where H represents the height (typically 1), and W signifies the width, representing the number of visible APs within the indoor environment. Each element in this tensor corresponds to the RSS measured by a specific AP at a particular RP. To convert these RSS fingerprint tensors into grayscale images, a mapping process is applied. This mapping function translates the RSS values into pixel intensities, ensuring that higher RSS values are represented with brighter pixels and lower RSS values with darker pixels. The resulting grayscale images have a shape of (N, H, W, C), where N denotes the RPs, H represents the height (usually 1), W signifies the width (number of visible APs), and C represents the number of channels (typically 1 for grayscale). This conversion preserves the spatial information of RSS across the indoor space, facilitating effective localization.

*C. Capsule Neural Network Architecture*

The capsule neural network is a pivotal component of the SENTINEL framework, comprising of five sub-components: the convolutional (CONV) layer, primary capsule (PC) layer, outer capsule (OC) layer, an agreement-based routing algorithm, and a majority voting layer. The enhanced capsule neural network in SENTINEL possesses several key differences from EDGELOC [28] which uses a simple capsule neural network: 1) Unlike [28], SENTINEL integrates a majority voting layer to enhance prediction output, 2) Unlike [28], SENTINEL is tailored specifically for processing grayscale fingerprint images, 3) [28] targets device heterogeneity only, whereas SENTINEL optimizes hyperparameters differently to simultaneously target mitigation of dynamic environment induced RSS fluctuations, device heterogeneity, and adversarial attacks, and 4) SENTINEL is pruned in the number of capsules (both PC and OC layers) and neurons within each capsule, resulting in a more lightweight deployment on resource-constrained edge devices than [28] while maintaining accuracy. We compare SENTINEL against EDGELOC [28] in Section V. In the rest of this section, we describe the various components of our SENTINEL capsule neural network.

- **Convolutional (CONV) layer:** The CONV layer captures spatial features within the grayscale fingerprint images. This layer employs convolutional filter kernels to extract distinctive patterns and features from the input images. Let us denote the grayscale RSS fingerprint image as *IM*, which has dimensions *(N, H, W, C)*. The convolutional layer consists of multiple filters kernels, denoted as *F*, which are applied to *IM*. The *F* slide across the entire *IM*, performing element-wise multiplications and summations, generating feature maps that highlight spatial features within the *IM*.

$$CON(p,q) = \sum_{i=0}^{H} \sum_{j=0}^{W} IM(p-i, q-j) * F(i,j) \quad (8)$$

In the equation above, $CON(p,q)$ denotes the feature at position $(p,q)$ in the CONV feature map and $F(i,j)$ represents the corresponding element of the filter kernel. $IM(p-i, q-j)$ represents the pixel value of *IM* at position $(p-i, q-j)$. The summation is performed over the height (*H*) and width (*W*) of *F*. During training, the network learns the optimal values of *F* through backpropagation. This process enables the CONV layer to automatically detect and extract relevant spatial features from the input RSS fingerprint images, providing meaningful representations that contribute to the overall accuracy of the localization process.

- **Primary capsule (PC) layer:** The PC layer receives the spatial features extracted by the CONV layer and serves as the next processing stage in the capsule neural network. A capsule is defined as a group of neurons, where each capsule within the PC layer generates a vector, referred to as the "activity vector". This vector captures both the magnitude (presence) and position of each feature in the RSS fingerprint. Unlike traditional neural networks (such as MLPs and CNNs) where neurons in subsequent layers are densely connected to all neurons in the preceding layer, the PC layer comprises of capsules, where each capsule corresponds to a specific spatial feature detected by the CONV layer. The activity vector ($u_{ij}$) for capsule *i* is obtained through a series of computations:

$$S_i = \sum_j V_{ij} * CON_j \quad (9)$$

$$u_{ij} = Squash(S_i) = \frac{||S_i||^2}{1+||S_i||^2} * \frac{S_i}{||S_i||} \quad (10)$$

In equation (9), $S_i$ represents the input for each capsule *i*, which is calculated as the weighted sum of outputs from the CONV layer using weight tensors ($V_{ij}$). These weight tensors determine the contribution of each feature from the CONV layer, enabling the PC layer to selectively focus on relevant spatial features. Subsequently, $S_i$ is squashed using a non-linear activation function known as the squash function. The squash function transforms $S_i$ into activity vectors $u_{ij}$, which represent the magnitude and position of the detected spatial features within the RSS fingerprint. This enables the PC layer to encode spatial relationships between features, enhancing the network's ability to capture meaningful representations of the indoor environment.

- **Outer capsule (OC) layer:** The OC layer performs classifications based on the activity vectors ($u_{ij}$), received from the PC layer. Each capsule in the OC layer corresponds to an RP class which determines the probability of the input fingerprint image belonging to that class. The classification process in the OC layer involves computing the agreement score between the $u_{ij}$ and the weights tensors ($W_{ij}$) associated with each capsule in the OC layer.

$$a_i = u_{ij} * W_{ij} \quad (11)$$
$$P_i = Softmax(a_i) \quad (12)$$

In equation (11), $a_i$ represents the agreement score for capsule *i*. The $W_{ij}$ contains the weight tensors associated with the connections between the PC and OC layers, determining the importance of each spatial feature for the classification of the corresponding RP class. In equation (12), $P_i$ denotes the predicted RP of capsule *i* after applying the *Softmax* function to $a_i$ from equation (11). This function assigns probabilities to each RP class based on $a_i$, facilitating the classification process.

- **Agreement-based routing algorithm:** The agreement-based routing algorithm plays a crucial role in refining the weight tensors ($W_{ij}$) between the PC and OC layers. After the OC layer receives activity vectors ($u_{ij}$) from the PC layer, the agreement scores ($a_i$) are computed using equation (11), representing the agreement between the $u_{ij}$ and $W_{ij}$ associated with each capsule in the OC layer. The goal of the routing algorithm is to iteratively adjust these weights tensors based on the $a_i$ achieved. The routing process involves several iterative steps, where $a_i$ are used to update the $W_{ij}$ in a way that maximizes agreement between the $a_i$ and the predicted RP classes. This iterative refinement enhances the network's ability to accurately classify input fingerprint images.
- **Majority voting layer:** The majority voting layer is the final component of the proposed capsule neural network. This layer aggregates the predictions ($P_i$) generated by the OC layer for each capsule. The majority voting mechanism aims to determine the final prediction by selecting the RP class with the highest number of aligned predictions from the capsules in the OC layer.

$$Prediction = Argmax\ (P_0, P_1, \ldots \ldots P_n) \quad (13)$$

In equation (13), $n$ represents the total number of RP classes. The $Argmax$ function selects the RP class with the highest probability as the final prediction. By ensuring that a majority of capsules agree on the final class, the majority voting layer reduces the impact of erroneous predictions from individual capsules. Additionally, the agreement-based routing algorithm ensures optimal predictions by refining the outputs between the PC and OC layers.

## V. EXPERIMENTS

### A. Experimental Setup

In this section, we describe our experimental setup, designed to evaluate the performance of our proposed SENTINEL framework in real-world scenarios. Our objective is to conduct comprehensive comparisons with state-of-the-art indoor localization frameworks, including CNNLOC [21], VITAL [27], EDGELOC [28], ADVLOC [31], and CALLOC [32], using simulated (FGSM, PGD, and MIM) and real-world *RSSRogueLoc* [35] data. To ensure the robustness of our evaluation, we embarked on an extensive data collection process. This involved gathering RSS fingerprints from diverse devices commonly available to the public, to capture performance across real-world scenarios. Data was collected during regular working hours, incorporating both dynamic and static occupants to reflect realistic conditions. Table I shows an overview of the real devices utilized in our experiments.

TABLE I: DEVICES USED TO COLLECT RSS FINGERPRINTS

| Device Name | Wi-Fi Chipset | Acronym | Year |
|---|---|---|---|
| BLU Vivo 8 | MediaTek Helio P10 | BLU | 2017 |
| Google Pixel 6a | Google Tensor G1 | GOOGLE | 2022 |
| HTC U11 | Qualcomm Snapdragon 835 | HTC | 2017 |
| Motorola Z2 | Qualcomm Snapdragon 835 | MOTO | 2017 |
| Nokia 7.1 | Qualcomm Snapdragon 636 | NOKIA | 2018 |
| OnePlus Nord 200 | Qualcomm Snapdragon 480 | ONEPLUS | 2021 |
| Xiaomi Redmi 10A | MediaTek Helio G88 | REDMI | 2022 |
| Samsung A14 | Samsung Exynos 850 | SAMSUNG | 2023 |

To ensure a comprehensive evaluation across diverse environmental conditions, we select building floorplans with varying factors such as path length, the number of visible APs, and environmental noise characteristics, as shown in Fig. 7. Our data collection strategy is designed to facilitate thorough training and testing of the SENTINEL framework. For each building floorplan, we allocate five fingerprints per RP for training and one fingerprint per RP, per device, and per building, for testing. Acknowledging the substantial effort required to gather a large volume of offline training data, we restrict the collection of offline data to a single device. To facilitate this, we designate the MOTO device as the primary training device. This approach aims to balance the need for sufficient training data with practical considerations regarding data collection efforts. All devices in Table I are used in the online phase during testing, again representing a practical scenario where developers of indoor localization frameworks need their solution to work across diverse user devices, even though their development phase targets a single device.

The SENTINEL framework is configured with specific architectural hyperparameters. The convolutional (CONV) layer is equipped with 32 filters and the PC layer comprises of 8 capsules with each capsule containing a dimension of 32 neurons. Furthermore, the OC layer contains capsules equal to the number of RP classes with a dimension of 32 neurons each, trained over 300 epochs using the Adam optimizer (learning rate = 0.001) and the sparse categorical cross-entropy loss function. The capsule neural network architecture results in a total of 2,117,687 trainable parameters, with a compact model size of 8.07 MB, facilitating low overhead deployment on most resource-constrained edge devices. Additionally, the SENTINEL framework incorporates an adversarial training mechanism aimed at enhancing its resilience against potential adversarial attacks. Adversarial samples are generated using the FGSM, PGD, and MIM approaches with $\epsilon$ set to 0.1 and $\varphi$ set to 100% (for training only). Each variant of our trained capsule neural network, augmented with adversarial samples, is denoted with a suffix. For instance, the model trained without adversarial samples is referred to as SENTINEL-NONE, while models trained with FGSM, PGD and MIM samples are labeled SENTINEL-FGSM, SENTINEL-PGD and SENTINEL-MIM, respectively. This enables us to further explore the performance of SENTINEL against various adversarial attack methods.

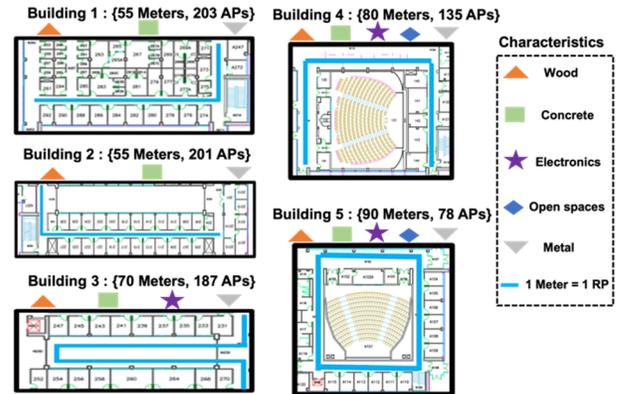

**Fig. 7.** Building floorplan layouts with varying path length, visible APs, and characteristics.

### B. Effects of Adversarial Training on Heterogeneity

In this section, we evaluate the performance of the SENTINEL framework under various adversarial training scenarios (FGSM, PGD, and MIM), separately. This experiment focuses on understanding the framework's efficacy across diverse building floorplans (with unique dynamic environments) and its response to device heterogeneity. In Fig. 8, we present heatmaps depicting the performance of the four SENTINEL variants: SENTINEL-NONE, SENTINEL-FGSM, SENTINEL-PGD, and SENTINEL-MIM. These models are individually trained on data collected exclusively from a single device (MOTO) and incorporate their respective adversarial training techniques. SENTINEL-NONE is trained without including any adversarial samples, providing a comparison of the effects of including adversarial training to the SENTINEL framework. Evaluation of these model variants are conducted using data acquired from all eight available devices across the five building floorplans, without any adversarial interference.

In Fig. 8, the X-axis of each heatmap represents the testing devices, while the Y-axis corresponds to the different buildings used for evaluation. Each cell within the heatmap indicates the average prediction error (in meters) across all RPs for a specific combination of test device and building floorplan. We observe differences in prediction errors across all the SENTINEL variants, due to the differences in adversarial training methods used. We note an increase in prediction errors when going from building 1 to 5, which can be attributed to increasing environmental dynamic causing higher variations in the selected building paths. For instance, building 1 exhibited low environmental noise, likely due to fewer people moving along the path during the testing. It also had relatively shorter path lengths, which overall resulted in lower prediction errors. In contrast, building 5 experienced higher environmental noise due to significantly more people moving along the path during the testing phase, and longer path lengths, leading to higher prediction errors. SENTINEL-FGSM consistently exhibits the lowest prediction errors, followed by SENTINEL-PGD, SENTINEL-NONE and SENTINEL-MIM. This trend suggests that while more advanced adversarial training methods like PGD and MIM may offer refined perturbations, they also introduce complexity and potential instability during training, leading to overfitting. The overfitting occurs because the adversarial samples generated by PGD and MIM involve multiple iterations of perturbations, making them more complex and causing feature mismatches between RP classes. As a result, the model may become overly specialized to these adversarial examples, reducing its ability to generalize well to unseen, real-world data. SENTINEL-FGSM however, stands out due to its balance between perturbation effectiveness and model stability. Its non-iterative nature allows for smaller, controlled perturbations, reducing the chances of a feature mismatch between legitimate and FGSM samples.

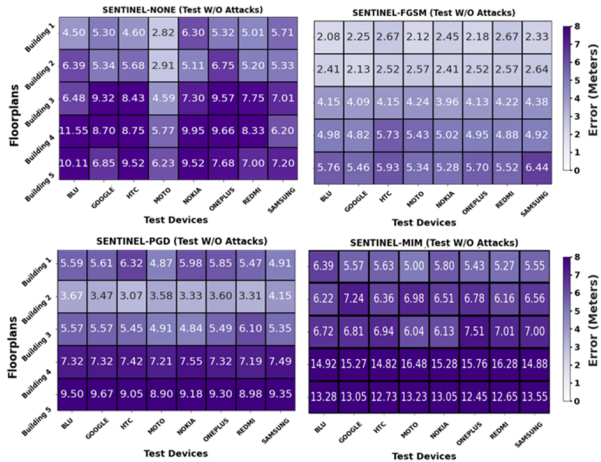

**Fig. 8.** Performance of the SENTINEL variants across different devices and building floorplans.

To more clearly illustrate the impact of device heterogeneity and assess the performance of the SENTINEL variants, we present Fig. 9. Here, the X-axis represents the testing devices, and the Y-axis denotes the prediction error in meters. Each bar represents the average prediction error per device across all building floorplans, with error bars included to indicate the range of errors observed per testing device, with the lower whisker representing the best case and the upper whisker representing the worst-case location error. In Fig. 9, we observe that the average error per testing device remains consistent for each SENTINEL variant. However, the SENTINEL-NONE variant exhibits the least consistency in prediction errors across the testing devices, with some devices showing higher errors while others show lower errors. Notably, the MOTO device consistently demonstrates significantly lower prediction errors compared to the other testing devices, indicating potential bias towards the MOTO device during inference. This suggests lower resilience to heterogeneity for the SENTINEL-NONE variant, as the MOTO device was used to train the SENTINEL model. Conversely, other SENTINEL variants show consistent prediction errors regardless of the training or testing devices used, indicating better heterogeneity resilience. Furthermore, incorporating adversarial training not only strengthens the robustness of the SENTINEL variants against adversarial attacks but also improves their resilience to heterogeneity. By subjecting the models to adversarial perturbations during training, the variants learn more generalized features, making them less sensitive to fluctuations from the testing devices. Particularly noteworthy is the performance of SENTINEL-FGSM, with up to 1.48× to 2.43× lower average and worst-case errors compared to the rest of the SENTINEL variants.

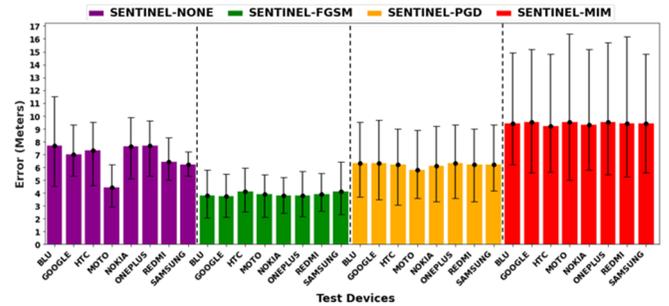

**Fig. 9.** Performance summary for SENTINEL variants.

*C. Evaluating the Impact of Varying Compromised APs (φ)*

In this section, we investigate the impact of varying the number of compromised APs (φ) in the testing phase, using different adversarial attack methods (FGSM, PGD, MIM), on the performance of the SENTINEL variants. To maintain consistency, we set the attack perturbation strength (ϵ) to 0.1, indicating 10% added perturbations per φ. In Fig. 10, the X-axis represents φ, ranging from 0 to 100, indicating 0% (no attacked APs) to 100% (all visible APs being attacked). The Y-axis denotes prediction errors measured in meters and the line plots illustrate the performance of each SENTINEL variant under the three adversarial attack methods. In Fig. 10, each marker (at different φ values) indicates the average prediction error across all testing devices and building floorplans.

We observe that as φ increases, the prediction errors for all SENTINEL variants also increase. However, there is a stabilization point observed at approximately φ = 50% for most variants methods (except SENTINEL-NONE, which lacks adversarial training), suggesting that the performance of the SENTINEL variants remains relatively unaffected when a significant portion of APs are compromised. This stabilization point indicates that the SENTINEL variants are resilient to attacks involving large numbers of compromised APs. Additionally, most variants demonstrate resilience against various adversarial attack methods (except SENTINEL-NONE), as evidenced by the almost flat line in prediction errors. Specifically, when subjected to the FGSM attack, the SENTINEL-FGSM model exhibits 1.90×, 2.35×, and 2.64× lower average errors compared to the SENTINEL-PGD, SENTINEL-NONE and SENTINEL-MIM models, respectively. Similarly, under the PGD attack, the SENTINEL-FGSM model demonstrates 1.69×, 2.75×, and 2.40× lower average errors compared to the SENTINEL-PGD, SENTINEL-NONE and SENTINEL-MIM models, respectively. Lastly, when influenced by the MIM attack, the SENTINEL-FGSM model shows 1.67×, 2.71×, and 2.15× lower average errors compared to the SENTINEL-PGD, SENTINEL-NONE and SENTINEL-MIM models, respectively. This superior performance of the SENTINEL-FGSM variant can be attributed to its FGSM-based adversarial training, which enhances the model's ability to recognize and adapt to adversarial perturbations during training. Hence, the SENTINEL-FGSM variant exhibits improved adversarial resilience and generalization capabilities compared to the other variants.

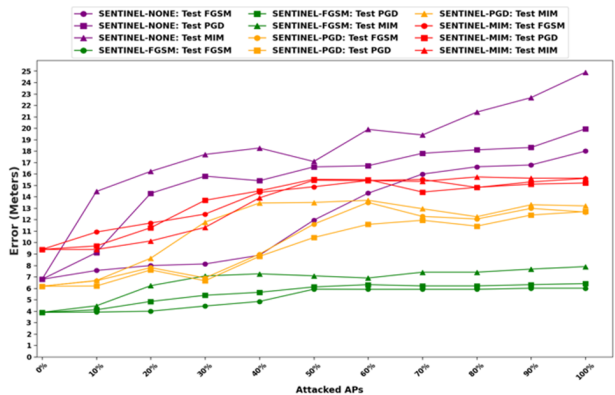

**Fig. 10.** Performance of the four SENTINEL variants on simulated adversarial attacks through varying φ.

### D. Evaluating the Impact of Varying Perturbations (ϵ)

In this section, we explore the impact of varying levels of perturbation strength (ϵ) in the testing phase on the performance of all SENTINEL variants. Our objective is to investigate how the prediction performance of each SENTINEL variant is affected by changes in ϵ, ranging from 0 (indicating no attack) to 0.5 (representing a 50% increase in added perturbations). In Fig. 11, the X-axis represents the varying levels of ϵ, while the Y-axis denotes the prediction error in meters. Each bar in the plot signifies the average prediction error across all testing devices, building floorplans, and φ values. Additionally, error bars are included to depict the range between the best (lower whisker) and worst-case (upper whisker) prediction errors. Our analysis reveals that as ϵ increases, there is a slight rise in prediction errors. However, we observe that all SENTINEL variants stabilize at approximately ϵ = 0.2 (except SENTINEL-NONE, lacking adversarial training). This suggests that regardless of the increase in perturbation strength, all SENTINEL models demonstrate consistent performance. Furthermore, we observe that the SENTINEL-FGSM variant consistently outperforms SENTINEL-PGD, SENTINEL-NONE and SENTINEL-MIM. On average, SENTINEL-FGSM demonstrates 1.48×, 2.81×, and 1.90× lower average prediction errors compared to SENTINEL-PGD, SENTINEL-NONE and SENTINEL-MIM, respectively. The superior performance of the SENTINEL-FGSM variant, even as ϵ increases during testing, can be attributed to the robustness gained through FGSM-based adversarial training. Although the model was trained with a fixed ϵ value, the adversarial training process encourages the model to capture underlying patterns in feature positions that are susceptible to adversarial attacks. This enables the model to generalize and adapt to perturbations even on varying ϵ. In contrast, other methods like PGD and MIM often induce significant perturbations in underlying features, leading to overfitting and reduced resilience during testing. The chosen epsilon range of 0 to 0.5 represents a practical attack range for indoor localization [39]. Additionally, the SENTINEL framework harnesses fingerprint image conversion techniques as discussed earlier to mitigate the impact of adversarial perturbations induced by larger epsilon values.

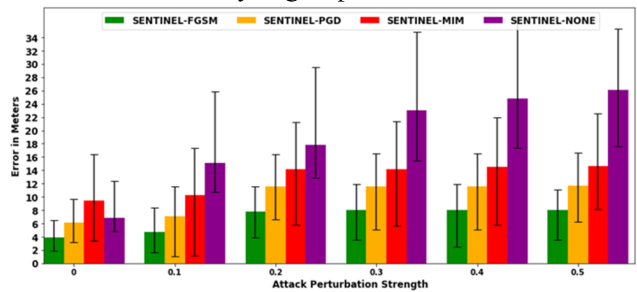

**Fig. 11.** Performance of the three SENTINEL variants on simulated adversarial attacks through varying ϵ.

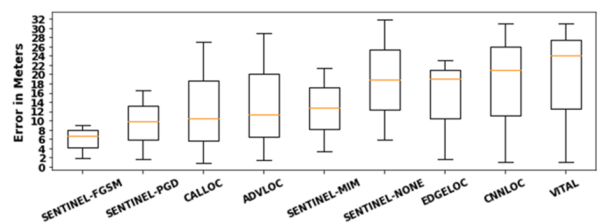

**Fig. 12.** Performance comparisons of all SENTINEL variants against state-of-the-art indoor localization frameworks.

### E. Comparison Against State-of-the-art Frameworks

In this section, we compare the performance of all SENTINEL variants against state-of-the-art indoor localization frameworks across various parameters including different devices, building floorplans, ϵ (ranging from 0 to 0.5), and φ (ranging from 0 to 100). Fig. 12 presents a box and whiskers plot, showcasing the comparison of the best case (lower whisker), worst case (upper whisker), and average (orange line) errors across all frameworks. This enhanced resilience can be attributed to the adversarial training and capsule neural network employed by the SENTINEL framework. The FGSM-based adversarial training introduces optimal adversarial features and feature dispositions (magnitude and positions), contrasting with other adversarial training methods that may lead to overfitting. The proposed capsule neural network treats each feature as a vector, effectively recognizing and capturing underlying patterns between the original (clean) and adversarial samples during training. This enables the SENTINEL-FGSM model to demonstrate lower prediction errors across various scenarios and metrics compared to the other frameworks. The SENTINEL-FGSM model demonstrates 1.47×, 1.55×, 1.68×, 1.91×, 2.82×, 2.83×, 3.13×, and 3.5× lower average errors compared to SENTINEL-PGD, CALLOC, ADVLOC, SENTINEL-MIM, SENTINEL-NONE, EDGELOC, CNNLOC, and VITAL, respectively. The SENTINEL-FGSM model also shows 1.83×, 3.0×, 3.2×, 2.37×, 2.55×, 3.4×, 3.4×, 3.4× lower worst-case errors compared to SENTINEL-PGD, CALLOC, ADVLOC, SENTINEL-MIM, EDGELOC, CNNLOC, SENTINEL-NONE, and VITAL, respectively.

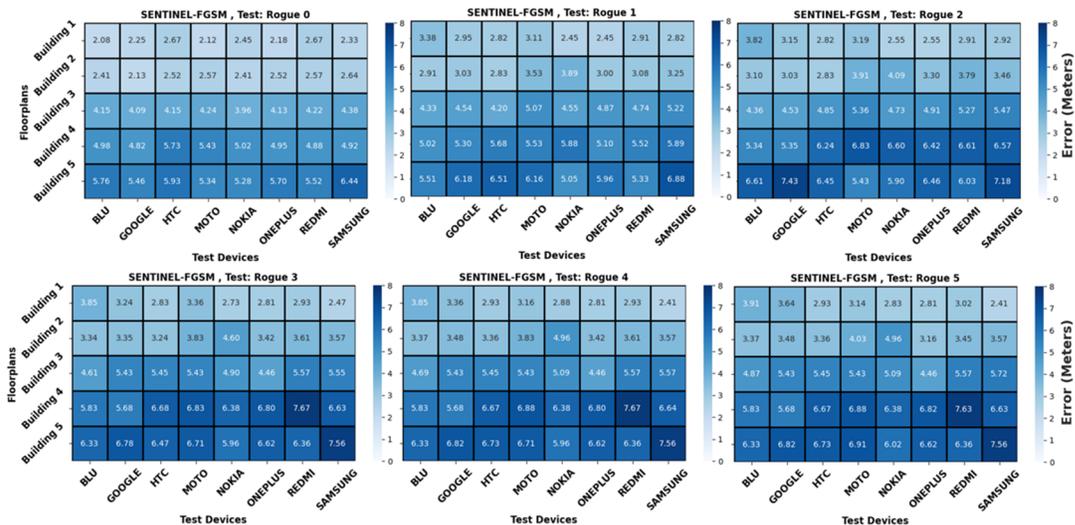

**Fig. 13.** Device-Floorplan wise performance of SENTINEL-FGSM on *RSSRogueLoc* dataset.

Additionally, recognizing the need for lightweight frameworks adaptable for resource-constrained edge devices, we analyze the parameter count and memory footprint of the various frameworks as shown in Table II. SENTINEL yields a relatively compact model size of 8.07 MB.

TABLE II: MODEL PARAMETERS, SIZE OF ALL FRAMEWORKS

| Framework | Total Parameters | Model Size |
|---|---|---|
| CALLOC | 652,390 | 2.48 MB |
| CNNLOC | 858,720 | 3.27 MB |
| ADVLOC | 1,746,752 | 6.99 MB |
| **SENTINEL** | **2,117,687** | **8.07 MB** |
| EDGELOC | 2,317,687 | 8.84 MB |
| VITAL | 2,347,006 | 8.95 MB |

*F. Evaluation on the New Real-World Rogue AP Attack Dataset*

In this section, we introduce a novel Wi-Fi RSS fingerprint dataset named *RSSRogueLoc* [35], designed to capture the detrimental effects of rogue APs for indoor localization systems. Unlike prior works which primarily rely on simulated adversarial attacks introduced by methods such as FGSM, PGD, and MIM, *RSSRogueLoc* delves into real-world adversarial scenarios, particularly those involving rogue APs. Building on the dataset outlined in Section V.A, *RSSRogueLoc* introduces a secondary testing dataset comprising up to five new devices configured as rogue APs (devices detailed in Table III), designed to execute evil twin attacks as discussed in Section III.A, where each rogue is configured to impact one legitimate AP. The *RSSRogueLoc* fingerprints were collected by incrementally introducing rogue APs across all RPs within each building floorplan. This sequential escalation started from Rogue 0, signifying the absence of all rogues, followed by Rogue 1 with one rogue per RP per floorplan, Rogue 2 with two rogues per RP per floorplan, Rogue 3 with three rogues per RP per floorplan, Rogue 4 with four rogues per RP per floorplan, and finally Rogue 5 with five rogues per RP per floorplan. The testing fingerprints were collected using all eight devices mentioned in Table I. This process unfolded over several weeks, to thoroughly capture the complexities of rogue AP configurations across numerous RPs and building floorplans. Each RP and floorplan underwent repeated scans with varying rogue configurations to ensure comprehensive coverage. This effort represents the first instance of compiling such a comprehensive dataset in the domain of indoor localization research. *To further facilitate research in this nascent field, we will open-source the RSSRogueLoc [35] dataset, to allow the indoor localization community to explore the impact of real-world rogue APs.*

TABLE III : ROGUE AP DEVICES USED IN *RSSROGUELOC*

| Device Name | Wi-Fi Chipset | Device Type |
|---|---|---|
| Samsung G991U | Samsung Exynos 2100 | Smartphone |
| Apple A2789 | Apple U2 | Laptop |
| HP 840 G6 | Intel Wi-Fi AX201 | Laptop |
| Vivo V2025 | Qualcomm Snapdragon 720G | Smartphone |
| HP 840 G10 | Intel Wi-Fi AX211 | Laptop |

We evaluate the performance of the best-performing SENTINEL variant (SENTINEL-FGSM), on the newly introduced *RSSRogueLoc* dataset. In Fig.13, the X-axis of each heatmap represents the testing devices, while the Y-axis corresponds to the building floorplans. Each cell within the heatmap indicates the average prediction error (in meters) across all RPs. The heatmaps present comprehensive results for each of the Rogue configurations tested on the SENTINEL-FGSM model. Notably, we observe minimal changes in errors even with an increase in the number of rogues in the respective building floorplans, suggesting that the SENTINEL-FGSM model effectively addresses adversarial attacks posed by rogue APs. To provide additional insights into the performance of all SENTINEL variants and state-of-the-art baseline frameworks on the *RSSRogueLoc* dataset, we present Fig. 14. This figure showcases a box and whisker plot, comparing the best-case (lower whisker), worst-case (upper whisker), and average (orange line) errors across all frameworks tested on the *RSSRogueLoc* dataset. The SENTINEL-FGSM model demonstrates 1.51×, 1.65×, 1.68×, 1.91×, 2.04×, 2.27×, 2.34×, and 2.80× lower average error compared to SENTINEL-PGD, CALLOC, ADVLOC, EDGELOC, SENTINEL-MIM, SENTINEL-NONE, CNNLOC, and VITAL, respectively. The SENTINEL-FGSM variant also shows a 1.62×, 1.69×, 1.81×, 2.08×, 2.01×, 2.28×, 2.48×, and 2.74× lower worst-case error compared to SENTINEL-PGD, CALLOC, ADVLOC, EDGELOC, SENTINEL-MIM, SENTINEL-NONE, CNNLOC, and VITAL, respectively.

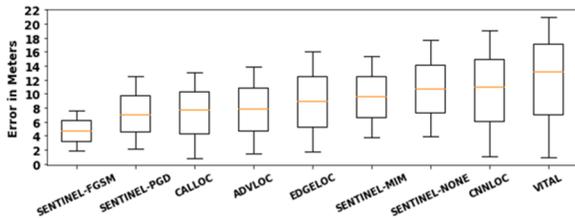

**Fig. 14.** Performance comparisons of all SENTINEL models against state-of-the-art on the *RSSRogueLoc* dataset.

## VI. CONCLUSION

The SENTINEL framework proposed in this work exhibits resilience against RSS fluctuations arising from environmental noise, edge device heterogeneity, and challenging adversarial attacks, due to its novel combination of adversarial training and modified capsule neural networks, while being relatively lightweight for edge device deployment. Through rigorous evaluation, we found that the SENTINEL-FGSM variant consistently achieves the lowest indoor localization errors, outperforming all baseline frameworks by 1.47× to 3.5× in average errors and 1.83× to 3.4× in worst-case errors on simulated adversarial attacks. Moreover, our introduction of the *RSSRogueLoc* dataset, designed to capture real-world effects of rogue APs (performing evil twin attacks in real-time), further highlights the superiority of the SENTINEL-FGSM variant. With 1.51× to 2.8× lower average errors and 1.63× to 2.74× lower worst-case errors compared to other state-of-the-art frameworks, our SENTINEL framework demonstrates its effectiveness in addressing practical challenges with security and reliability of indoor localization applications.

## REFERENCES

[1] S. Tiku, S. Pasricha. "An Overview of Indoor Localization Techniques." *IEEE CEM*, 2017
[2] D. Lymberopoulos, J. Liu. "The microsoft indoor localization competition: Experiences" *IEEE SPM*, 2017
[3] A. Petrenko, A. Sizo, W. Qian, AD. Knowles, A. Tavassolian, K. Stanley, S. Bell. "Exploring mobility indoors: an application of sensor-based and GIS systems." *Transactions in GIS*, 2014
[4] WK. Zegeye, SB. Amsalu, F. Moazzami. "WiFi RSS fingerprinting indoor localization for mobile devices." *IEEE UEMCON*, 2016
[5] M. Shu, G. Chen, Z. Zhang. "Efficient image-based indoor localization with MEMS aid on the mobile device." *ISPRS*, 2022
[6] X. Tian, R. Shen, D. Liu, Y. Wen, X. Wang. "Performance analysis of RSS fingerprinting based indoor localization." *IEEE TMC, 2016*
[7] B. Yang, L. Guo, R. Guo, M. Zhao, T. Zhao. "A novel trilateration algorithm for RSSI-based indoor localization." *IEEE Sensors 2020*.
[8] X. Hou, T. Arslan, A. Juri, F. Wang. "Indoor localization for bluetooth low energy devices using weighted off-set triangulation algorithm." *ION GNSS,* 2016
[9] F. Alhomayani, MH. Mahoor. "Deep learning methods for fingerprint-based indoor positioning: A review." *Journal of Location Based Services,* 2020
[10] N. Singh, S. Choe, R. Punmiya. "Machine learning based indoor localization using Wi-Fi RSSI fingerprints: An overview." *IEEE access,* 2021
[11] D. Duong, Y. Xu, K. David. "The influence of fast fading and device heterogeneity on wi-fi fingerprinting." *IEEE VTC*, 2018
[12] I. Alshami, N. Ahmad, S. Sahibuddin, "RSS Certainty: An Efficient Solution for RSS Variation due to Device Heterogeneity in WLAN Fingerprinting-based Indoor Positioning System." *PICICT*, 2021
[13] I. J. Goodfellow, J. Shlens, C. Szegedy. "Explaining and harnessing adversarial examples. " *arXiv:1412.6572*, 2014
[14] C. Szegedy, W. Liu, Y. Jia, P. Sermanet, S. Reed, D. Anguelov, D. Erhan, V. Vanhoucke, A. Rabinovich. "Going deeper with convolutions. " *IEEE CVPR*, 2015
[15] W. Xue, X. Hua, Q. Li, W. Qiu, X. Peng. "Improved clustering algorithm of neighboring reference points based on KNN for indoor localization." *IEEE UPINLBS*, 2018
[16] J. Jadidi, M. Patel, JV. Miro. "Gaussian processes online observation classification for RSSI-based low-cost indoor positioning systems." *IEEE ICRA,* 2017
[17] AB. Adege, HP. Lin, GB. Tarekegn, SS. Jeng. "Applying deep neural network (DNN) for robust indoor localization in multi-building environment." *Applied Sciences,* 2018
[18] M. Dakkak, B. Daachi, A. Nakib, P. Siarry. "Multi-layer perceptron neural network and nearest neighbor approaches for indoor localization." *IEEE SMC*, 2014
[19] W. Liu, H. Chen, Q. Cheng, "LC-DNN: Local Connection Based Deep Neural Network for Indoor Localization With CSI*." IEEE Access*, 2020
[20] Y. Kim, H. Shin, Y. Chon, H. Cha. "Smartphone-based Wi-Fi tracking system exploiting the RSS peak to overcome the RSS variance problem." *Pervasive and Mobile Computing*, 2013
[21] X. Song, X. Fan, X. He, C. Xiang, Z. Wang. "Cnnloc: Deep-learning based indoor localization with wifi fingerprinting." *IEEE SUI*, 2019
[22] D. Gufran, S. Tiku, S. Pasricha. "SANGRIA: Stacked autoencoder neural networks with gradient boosting for indoor localization." *IEEE ESL,* 2023
[23] S. Tiku, D. Gufran, S. Pasricha. "Multi-head attention neural network for smartphone invariant indoor localization." *IEEE IPIN*, 2022
[24] Z. Zhang, H. Du, S. Choi. "Tips: Transformer based indoor positioning system using both csi and doa of wifi signal." *IEEE Access,* 2022
[25] G. Elsayed, P. Ramachandran, J. Shlens, S. Kornblith. "Revisiting spatial invariance with low-rank local connectivity." *PMLR,* 2020
[26] O. Parkhi, A. Ved, A. Zisserman. "Deep face recognition." *BMVC*, 2015
[27] D. Gufran, S. Tiku, S. Pasricha. "VITAL: Vision transformer neural networks for accurate smartphone heterogeneity resilient indoor localization." *IEEE DAC,* 2023
[28] Q. Ye, H. Bie, KC. Li, X. Fan, G. Fang. "Edgeloc: A robust and real-time localization system toward heterogeneous iot devices." *IOT,* 2021
[29] A. Shafahi, M. Najibi, MA. Ghiasi, Z. Xu, LS. Davis, G. Taylor, T. Goldstein. "Adversarial training for free!." *NIPS*, 2019
[30] A. Kurakin, I. Goodfellow, S. Bengio. "Adversarial machine learning at scale." *arXiv:1611.01236*, 2016
[31] X. Wang, X. Wang, S. Mao, J. Zhang, SCG. Periaswamy, J. Patton. "Adversarial deep learning for indoor localization with channel state information tensors. " *IEEE IOT,* 2022
[32] D. Gufran, S. Pasricha. "CALLOC: Curriculum adversarial learning for secure and robust indoor localization." *arXiv:2311.06361,* 2023
[33] S. Finlayson, HW. Chung, I. Kohane, A. Beam. "Adversarial attacks against medical deep learning systems." *arXiv:1804.05296*, 2018
[34] Y. Dong, F. Liao, T. Pang, H. Su, J. Zhu, X. Hu, and J. Li. "Boosting adversarial attacks with momentum." *IEEE CVPR*, 2018
[35] EPIC-CSU "Heterogeneous RSSI Indoor Navigation," GitHub, 2023, [Online] https://github.com/EPIC-CSU/heterogeneous-rssi-indoor-nav
[36] Q. Lu, H. Qu, Y. Zhuang, X. Lin, Y. Zhu, Y. Liu, "A Passive Client-based Approach to Detect Evil Twin Attacks." *IEEE ICESS*, 2017

[37] A. Mallik, "Man-in-the-middle-attack: Understanding in simple words." *Cyberspace*, 2019
[38] S. Sabour, N. Frosst, GE. Hinton, "Dynamic routing between capsules." *NeurIPS*, 2017
[39] M. Patil, X. Wang, S. Mao, "Adversarial attacks on deep learning-based floor classification and indoor localization" *ACM WSML* 2021
[40] D. Gufran, S. Pasricha. "FedHIL: Heterogeneity resilient federated learning for robust indoor localization with mobile devices" *ACM TECS,* 2023
[41] D. Gufran, S. Tiku, S. Pasricha. "STELLAR: Siamese multiheaded attention neural networks for overcoming temporal variations and device heterogeneity with indoor localization" *IEEE ISPIN,* 2023
[42] S. Tiku, D. Gufran, S. Pasricha. " Smartphone invariant indoor localization using multi-head attention neural network" *Machine Learning for Indoor Localization and Navigation*, 2023
[43] D. Gufran, S. Tiku, S. Pasricha. " Heterogeneous device resilient indoor localization using vision transformer neural networks" *Machine Learning for Indoor Localization and Navigation*, 2023

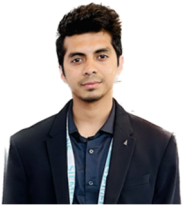
**Danish Gufran** (Student Member, IEEE), received his Bachelor of Engineering (B.E.) degree in Electronics and Instrumentation Engineering from Visvesvaraya Technological University, India, in 2020. He is currently pursuing his Ph.D. in Computer Engineering at Colorado State University, Fort Collins. His research focuses on designing innovative low-power deep machine learning algorithms to develop energy-efficient models suitable for edge and IoT devices. Additionally, his work extends to implementing federated learning systems and adversarial security models for indoor localization. His contributions have been published in various IEEE and ACM conferences and journals. He is the recipient of a best paper award and has co-authored multiple book chapters. He is a student member of IEEE and ACM.

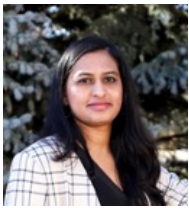
**Pooja Anandathirtha** received her Bachelor of Engineering (B.E.) degree in Telecommunication Engineering from Visvesvaraya Technological University, India. She is currently pursuing her Master of Science (M.S) in Electrical Engineering from Colorado State University, Fort Collins. Her research focuses on designing machine learning algorithms for indoor localization systems using embedded and IoT devices.

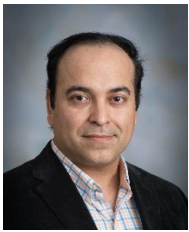
**Sudeep Pasricha** (Fellow, IEEE), received his Ph.D. in Computer Science from the University of California, Irvine in 2008. He is currently a Walter Scott Jr. College of Engineering Professor in the Department of Electrical and Computer Engineering, at Colorado State University. His research focuses on the design of innovative software algorithms, hardware architectures, and hardware-software co-design techniques for energy-efficient, fault-tolerant, real-time, and secure computing. He has co-authored five books and published more than 300 research articles in peer-reviewed journals and conferences that have received 17 Best Paper Awards and Nominations at various IEEE and ACM conferences. He has served as General Chair and Program Committee Chair for multiple IEEE and ACM conferences and served in the Editorial board of multiple IEEE and ACM journals. He is a Fellow of the IEEE, Fellow of the AIAA, Distinguished Member of the ACM, and an ACM Distinguished Speaker.